\documentclass[oneside,english]{amsart}
\usepackage[T1]{fontenc}
\usepackage[utf8]{luainputenc}
\usepackage{babel}
\usepackage{amsthm}
\usepackage{amssymb}
\usepackage[unicode=true,pdfusetitle,
 bookmarks=true,bookmarksnumbered=false,bookmarksopen=false,
 breaklinks=false,pdfborder={0 0 1},backref=false,colorlinks=false]
 {hyperref}

\makeatletter
\numberwithin{equation}{section}
\numberwithin{figure}{section}
\theoremstyle{plain}
\newtheorem{thm}{\protect\theoremname}
  \theoremstyle{definition}
  \newtheorem{defn}[thm]{\protect\definitionname}
  \theoremstyle{plain}
  \newtheorem{cor}[thm]{\protect\corollaryname}
  \theoremstyle{remark}
  \newtheorem{note}[thm]{\protect\notename}

\makeatother

  \providecommand{\corollaryname}{Corollary}
  \providecommand{\definitionname}{Definition}
  \providecommand{\notename}{Note}
\providecommand{\theoremname}{Theorem}

\begin{document}

\title{Small Jump with Negation-UTM Trampoline}

\author{Koji KOBAYASHI}

\maketitle

\section{Introduction}

This paper divide some complexity class by using fixpoint and fixpointless
area of Decidable Universal Turing Machine (UTM). Decidable Deterministic
Turing Machine (DTM) have fixpointless combinator that add no extra
resources (like Negation), but UTM makes some fixpoint in the combinator.
This means that we can jump out of the fixpointless combinator system
by making more complex problem from diagonalisation argument of UTM.

As a concrete example, we proof L is not P . We can make Polynomial
time UTM that emulate all Logarithm space DTM (LDTM). LDTM set close
under Negation, therefore UTM does not close under LDTM set. (We can
proof this theorem like halting problem and time/space hierarchy theorem,
and also we can extend this proof to divide time/space limited DTM
set.) In the same way, we proof P is not NP. These are new hierarchy
that use UTM and Negation.

\section{L is not P}
\begin{defn}
\label{def:Turing-Machine}``$DTM$'' is defined as Decidable Deterministic
Turing Machine set. ``$LDTM$'' is defined as logarithmic space
$DTM$. ``$pDTM$'' is defined as polynomial time $DTM$. ``$\bigcirc DTM$''
is defined as $DTM$ that some resource (time, space) limited.

``$UTM$'' is defined as Universal Turing Machine set. ``$UTM\left(C\right)$''
is defined as minimum $UTM$ that can emulate all $M\in C$. $\left\langle M\right\rangle $
is defined as code number of a $M\in DTM$ that $U\in UTM$ emulate.
That is, $\forall w\left[U\left(\left\langle M\right\rangle ,w\right)=M\left(w\right)\right]$
and $U\left(\left\langle M\right\rangle \right)=M$.

``$Negate\left(C\right)$'' is defined as minimum Negation system
that include $C$. That is, $\forall C\left[\left(C\subset Negate\left(C\right)\right)\wedge\left(\forall c\in Negate\left(C\right)\left[\neg c\in Negate\left(C\right)\right]\right)\right]$.\end{defn}
\begin{thm}
\label{thm:DTM-Negation}$\forall r\in\bigcirc DTM\left(\neg r\in\bigcirc DTM\right)$\end{thm}
\begin{proof}
It is trivial from DTM structure.

If DTM is

$M=\left(Q,\Sigma,\Gamma,\delta,q_{0},q_{1},q_{2}\right)$

then this dual machine

$\overline{M}=\left(Q,\Sigma,\Gamma,\delta,q_{0},q_{2},q_{1}\right)$

compute $\neg M$ without extra resources.

Therefore negation of $\bigcirc DTM$ is also in $\bigcirc DTM$.\end{proof}
\begin{thm}
\label{thm:LUTM}$\exists U\in UTM\left(LDTM\right)\left[U\in pDTM\right]$\end{thm}
\begin{proof}
It is trivial because some $U^{\prime}\in UTM$ can emulate all $LDTM$
in polynomial time. Therefore, we can make $U\in pDTM$ by limiting
at polynomial time (if $U^{\prime}$ compute over polynomial time,
$U$ reject these input).\end{proof}
\begin{thm}
\label{thm:L-is-not-P}$L\subsetneq P$\end{thm}
\begin{proof}
We can proof this theorem like halting problem and time/space hierarchy
theorem.

Because of

$\forall U\in UTM\left(LDTM\right),M\in LDTM\left[U\left(\left\langle M\right\rangle \right)=M\right]$
\ref{def:Turing-Machine}

all $M\in LDTM$ have index $\left\langle M\right\rangle $. Therefore
we can make $H$ which is diagonalization of $U$.

$H\left(\left\langle M\right\rangle \right)=U\left(\left\langle M\right\rangle ,\left\langle M\right\rangle \right)$

$\begin{array}{ccccccc}
 &  & \left\langle M_{0}\right\rangle  & \left\langle M_{1}\right\rangle  & \left\langle M_{2}\right\rangle  & \left\langle M_{3}\right\rangle  & \cdots\\
M_{0} & =\{ & \underline{\top} & \top & \bot & \top & \cdots\\
M_{1} & =\{ & \bot & \underline{\bot} & \top & \bot & \cdots\\
M_{2} & =\{ & \bot & \bot & \underline{\bot} & \top & \cdots\\
M_{3} & =\{ & \top & \top & \bot & \underline{\bot} & \cdots\\
\vdots &  & \vdots & \vdots & \vdots & \vdots\\
H & =\{ & \top & \bot & \bot & \bot & \cdots
\end{array}$

$H\in pDTM$ because $U\in pDTM$ \ref{thm:DTM-Negation} and $H$
input size is at least half of $U$.

Mentioned above \ref{thm:DTM-Negation},

$\forall r\in LDTM\left(\neg r\in LDTM\right)$

we can make $G$ which is Negation of diagonalization.

$G\left(\left\langle M\right\rangle \right)=\neg H\left(\left\langle M\right\rangle \right)=\neg U\left(\left\langle M\right\rangle ,\left\langle M\right\rangle \right)$ 

$\begin{array}{ccccccc}
 &  & \left\langle M_{0}\right\rangle  & \left\langle M_{1}\right\rangle  & \left\langle M_{2}\right\rangle  & \left\langle M_{3}\right\rangle  & \cdots\\
M_{0} & =\{ & \underline{\top} & \top & \bot & \top & \cdots\\
M_{1} & =\{ & \bot & \underline{\bot} & \top & \bot & \cdots\\
M_{2} & =\{ & \bot & \bot & \underline{\bot} & \top & \cdots\\
M_{3} & =\{ & \top & \top & \bot & \underline{\bot} & \cdots\\
\vdots &  & \vdots & \vdots & \vdots & \vdots\\
H & =\{ & \top & \bot & \bot & \bot & \cdots\\
G & =\{ & \bot & \top & \top & \top & \cdots
\end{array}$

$G\notin LDTM$ because $\forall M\in LDTM\left[G\left(\left\langle M\right\rangle \right)\neq M\left(\left\langle M\right\rangle \right)\right]$.
On the other hand, $G\in pDTM$ because $H\in pDTM$.

Therefore, $G\in pDTM\left(G\notin LDTM\right)$ and $L\subsetneq P$.
\end{proof}
We can expand above result to general DTM.
\begin{thm}
\label{thm:Negation-UTM-Jump} $\forall CC\subset DTM\left[Negate\left(UTM\left(Negate\left(CC\right)\right)\right)\nsubseteq Negate\left(CC\right)\right]$\end{thm}
\begin{proof}
We omit the proof because this proof is same as previous.\end{proof}
\begin{cor}
\label{cor:UTM and DTM}$Negate\left(UTM\left(\bigcirc DTM\right)\right)\nsubseteq\bigcirc DTM$
\end{cor}

\section{P is not NP}
\begin{thm}
\label{thm:pUTM}$\exists U\in UTM\left(pDTM\right)\left[U\in pNTM\right]$\end{thm}
\begin{proof}
We can make some oracle TM which oracle emulate transition function.

$np^{p}\mid np\in NP,p\in P$ 

$p\left(\left\langle t\right\rangle ,w\right)=t\left(w\right)$

$t\in P$: transition function

$\left\langle t\right\rangle $:code number of transition function
$t$

$w$: $t$'s input (state and symbol)

$p\left(\left\langle t\right\rangle ,w\right)$ accept if and only
if $t\left(w\right)$ accept and output $t\left(w\right)$.

Oracle TM $np^{p}\in pNTM$ and $np^{p}$ can emulate all $pDTM$.
Therefore $np^{p}\in UTM\left(pDTM\right)$.\end{proof}
\begin{note}
$np^{p}$ can change transition function more flexible than $pDTM$
in less time. In fact, $np^{p}$ can increase transition function
with logarithm time (by computing these transition functions in parallel).
Therefore, $np^{p}$ have more chance to compute more complex problems.\end{note}
\begin{thm}
\label{thm:P-is-not-NP}$P\subsetneq NP$\end{thm}
\begin{proof}
(Proof by contradiction.) Assume to the contrary that $P=NP$.

$P=NP$ means that $P=NP=coNP=PH$, therefore $NP$ close under Negation.

In the same way as mentioned above \ref{thm:L-is-not-P}, we get $P\subsetneq NP$.
This result contradicting assumption $P=NP$.

Therefore $P\subsetneq NP$.\end{proof}
\begin{cor}
\label{thm:P-is-not-coNP} $P\subsetneq coNP$
\end{cor}

\section{Trampoline Hierarchy between Negation and UTM}

This result shows that we can jump over border of asymptotic analysis
by using Negation (fixpointless combinator) and UTM (fixpoint creator).
Therefore, combination of UTM and Negation make new complexity class.
That is, there are some Hierarchy of UTM and Negation.


\begin{thebibliography}{1}
\appendix
\bibitem{Book1}Michael Sipser, (translation) Kazuo OHTA, Keisuke
TANAKA, Masayuki ABE, Hiroki UEDA, Atsushi FUJIOKA, Osamu WATANABE,
``Introduction to the Theory of COMPUTATION Second Edition (Japanese
version)'', 2008\end{thebibliography}
\end{document}